\documentclass{aa}
\usepackage{natbib}
\usepackage{graphicx}
\usepackage{txfonts}
\usepackage[dvips]{color}
\newcommand{\aflux}{A$_\mathrm{flux}$}
\newcommand{\hi}{H\,{\sc i}}
\newcommand{\hii}{H\,{\sc ii}}
\newcommand{\km}{km\,s$^{-1}$}
\newcommand{\lfir}{L$_\mathrm{FIR}$}
\newcommand{\prim}{$^{\prime}$}

\newcommand{\msolar}{M$_{\odot}$}
\newcommand{\lsolar}{L$_{\odot}$}
\newcommand{\LB}{L$_{B}$}

\newcommand{\degree}{$^{\circ}$}
\newcommand{\halpha}{H${\alpha}$}
\definecolor{gold}{rgb}{0.85,.66,0}
\newcommand{\acm}{atoms\,cm$^{-2}$}
\begin{document}

   \title{H{\sc i} asymmetry in the isolated  galaxy CIG\,85 (UGC 1547)}

   \subtitle{}

   \author{C. Sengupta
          \inst{1,2}
           \fnmsep\thanks{},
           T. C. Scott
           \inst{1},
           L. Verdes Montenegro
           \inst{1},
           A. Bosma
           \inst{3},
           S. Verley
           \inst{4},
           J. M. Vilchez
           \inst{1},
           A. Durbala
           \inst{5},
         \textcolor{black}{  M. Fern\'{a}ndez Lorenzo
           \inst{1},}
           D. Espada
           \inst{6},
           M. S. Yun
           \inst{7},
           E. Athanassoula
           \inst{3},
           J. Sulentic
           \inst{1},
           \and
           A. Portas
           \inst{1}
          }

   \institute{\textcolor{black}{1} Instituto de Astrof\'{i}sica de Andaluc\'{i}a (IAA/CSIC), Glorieta de la Astronomia, s/n, 18080 Granada, Spain. \\
\textcolor{black}{2} Calar--Alto Observatory, Centro Astron\'omico Hispano Alem\'an, C/Jes\'us Durb\'an Rem\'on, 2-2 04004 Almeria, Spain\\
\textcolor{black}{3} Aix Marseille Universite,CNRS, LAM (Laboratoire d'Astrophysique de Marseille) UMR 7326, 13388, Marseille, France\\
\textcolor{black}{4} Dept. de F\'isica Te\'orica y del Cosmos, Facultad de Ciencias, Universidad de Granada, Spain\\
\textcolor{black}{5} University of Wisconsin -Stevens Point, Department of Physics and Astronomy, 2001 Fourth Avenue, Stevens Point, WI 54481-1957 \\
\textcolor{black}{6} National Astronomical Observatory of Japan (NAOJ),
2-21-1 Osawa, Mitaka, Tokyo 181-8588, Japan\\
\textcolor{black}{7} Department of Astronomy, University of Massachusetts, 710 North Pleasant Street, Amherst, MA 01003, USA\\
              \email{sengupta@iaa.es}}

   \date{}

 
  \abstract
  {We present the results from the Giant Metrewave Radio Telescope (GMRT) interferometric \hi\ and 20 cm radio continuum  observations of CIG\,85, an isolated asymmetric galaxy from the AMIGA (Analysis
of the Interstellar Medium of Isolated
GAlaxies (http://amiga.iaa.es) sample. }
  {Despite being an isolated galaxy, CIG\,85 showed an appreciable optical and H{\sc i} spectral asymmetry and therefore was an excellant candidate for resolved H{\sc i} studies to understand the reasons giving rise to asymmetries in isolated galaxies.}
  {The galaxy was imaged in \hi\ and 20 cm radio continuum using the \textcolor{black}{GMRT. For a detailed discussion of the results
we also made use of multi--wavelength data from archival SDSS, GALEX and  H$\alpha$ imaging.  }}
   {We find the \hi\ in CIG 85 to have a clumpy, asymmetric distribution which in the NW part is correlated with optical tail like features, but the \hi\ velocity field displays a relatively regular rotation pattern.  Evaluating all the observational evidence, we come to a conclusion that CIG\,85 is most likely a case of a disturbed spiral galaxy which now appears to have the morphology of an irregular galaxy. \textcolor{black}{Althoug}h it is currently isolated from major companions, the observational evidence is consistent with \hi\  asymmetries, a highly disturbed optical disk and recent increase in star formation having been caused by a minor  merger, remnants of which are  now projected in front of the optical disk. If this is correct, the companion will be fully  accreted by CIG\,85 in the near future.}
 {}


\titlerunning{CIG\,85 \hi\ asymmetries}
\authorrunning{Sengupta et al.}

   \maketitle
%



\section{Introduction}
The evolution of a galaxy and its properties at z = 0  depend \textcolor{black}{on
 both} internal processes and \textcolor{black}{on} its environment. To quantify the impact of
different environments (nurture) on a galaxy's morphology,
structure, nuclear activity, or star formation (SF) properties, amongst
others, requires
a well defined and  statistically significant sample of  minimally
perturbed galaxies (pure nature). The AMIGA (Analysis
of the Interstellar Medium of Isolated
GAlaxies (http://amiga.iaa.es) project  \citep{vm05} provides
such a sample. The AMIGA catalogue is a refinement of the Catalogue of
Isolated Galaxies \citep[CIG;][]{kara73}. Galaxies in the AMIGA sample
have remained free of  major tidal interactions for the last $\sim$3 Gyr
\citep{vm05}. Quantification of the strength of tidal interactions
with neighbouring minor companions and the local number density is
available for all  of the AMIGA sample galaxies (Verley et
al. 2007a,b). \nocite{verley07a} \nocite{verley07b} The AMIGA project
has clearly established that the most isolated galaxies have different
physical properties, even compared to galaxies in what are generally considered
field samples, in terms of their optical morphology (e.g. asymmetry,
concentration), \lfir,  radio--far infrared correlation, molecular gas content, \textcolor{black}{fraction} of 
active galactic nuclei and
\hi\ spectra asymmetry \citep{lisen07,2008A&A...485..475L,2008A&A...486...73S,durbala08,espada11,2011A&A...534A.102L}. 

Despite having lower rates of \hi\ spectral and optical morphological asymmetry than galaxies in denser environments \citep[][]{durbala08,espada11}, some  galaxies in the AMIGA sample show an appreciable asymmetry. \cite{espada11} studied the \hi\ profiles of a sample of 166 AMIGA galaxies using an \hi\ asymmetry parameter \aflux,  defined as the ratio of the \hi\ flux between the receding and approaching sides of the single dish spectrum. They found the distribution of this parameter is well described by \textcolor{black}{the right half of a } Gaussian distribution, with only 2\% of the sample having an asymmetry parameter in excess of 3$\sigma$ (i.e. \aflux $>$ 1.39), meaning a 39\% excess of flux in one half of the spectrum.  They also noted that the fraction of asymmetric \hi\ profiles was smaller in the AMIGA sample than in any other samples  in the literature.  They found no  correlation between the \hi\ asymmetry parameter and minor companions, measured either as the tidal force (one-on-one interactions) or in terms of the number density of neighbouring galaxies. In contrast, field galaxy samples deviate from a Gaussian distribution and have higher (10-20\%) rates of asymmetric galaxies \citep{espada11}. 


It is well known that environment affects 
galaxy evolution: tidal interactions can stretch and deform both the stellar
and gas disks, ram pressure is able to deform and strip \hi\ disks
\citep{vgork04}, and major mergers can destroy the structure of the
disks \citep{struck99}.  The presence of asymmetric \hi\ profiles in
isolated galaxies and the Gaussian distribution of their
\aflux\ parameters \citep{espada11} implies that  
process(es) other than major interactions and
mergers are operating to  maintain  long lived or frequent perturbations in
isolated late--type galaxies. 


 For isolated galaxies 
  with \hi\ and stellar asymmetries, several 
 secular perturbation processes  have been proposed as the cause\textcolor{black}{:}  accretion of cold gas from the surrounding environment, intermittent accretion of satellite galaxies, and  internal bar, disk or retarded SF driven 
perturbations together with  their associated star formation \citep{bergvall95,bournaud02,bournaud05,sancisi08}. To date   very few \textcolor{black}{ detailed} observational  studies have been carried out   to determine \textcolor{black}{the} causes \textcolor{black}{of} such asymmetries in isolated galaxies.

\begin{figure}
\centering
\includegraphics[scale=0.54]{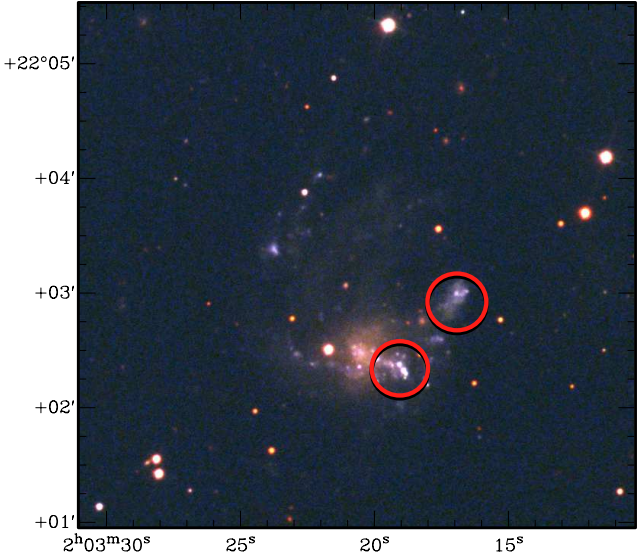}  
\caption{CIG\, 85 SDSS \textcolor{black}{\textit{u, g, r}}  image. The red circles indicate the positions of two possible dwarf galaxies 020036+21480-b  (lower circle)  and 020036+21480-c (top circle) reported by \cite{woods06}. }
\label{sdss}
\end{figure}

We are therefore carrying out a programme of \textcolor{black}{spatially} resolved \hi\ studies for a sample of AMIGA galaxies to \textcolor{black}{determine the} mechanisms that predominantly give rise to asymmetries in isolated galaxies \textcolor{black}{  \citep{espada05,espada11,portas}}. As part of that study  we present here \textcolor{black}{Giant Metrewave Radio Telescope (GMRT)} \hi\ observations of CIG\,85 (UGC 1547), \textcolor{black}{ (Figure \ref{sdss})} which is an  asymmetric AMIGA late--type \textcolor{black}{galaxy} with \aflux\ = 1.27 $\pm$ 0.01.

\textcolor{black}{Basic parameters for  CIG\,85, which is an optically irregular   galaxy, are given in Table \ref{table_a}.}  According to the AMIGA isolation criteria discussed in \cite{verley07b}, a galaxy needs to have the density and tidal force parameters of $\eta_k$ $\le$ 2.4 and Q$_{kar}$ $\le$ -2.0   respectively, to be included in the isolated sample of galaxies. These parameters for CIG\,85 are 0.984 and -3.74 respectively, making it highly isolated even \textcolor{black}{within} the AMIGA sample. In addition we have searched the literature for companions.  \cite{arp85}  suggest this galaxy is part of a small group with three neighbours, \textcolor{black}{NGC 772 (2474 \km), NGC 770 (2458 \km) and UGC 1551 (2670 \km). Of these, UGC 1551 is the nearest, projected 122$^{\prime}$ (1.2 Mpc) from CIG\,85. In addition to the three catalogued group members \citep{arp85}, a search of the NASA Extragalactic database (NED) revealed two nearer neighbours with similar optical magnitude, UGC\,1538 (2835 \km), 103$^{\prime}$ (1 Mpc) away and the closest UGC\,1490 (3079 \km), 60$^{\prime}$ (0.6 Mpc) away. Using 200 \km\ as the separation velocity, the time taken to cover 1 Mpc would be $\sim$ 3 Gyrs.} This is more than twice the dynamical time scale of CIG\,85. As it is generally thought that stellar and \hi\ asymmetries caused by interactions dissipate within the time it takes for the galaxy to complete a single rotation, it may safely be assumed that a major interaction with known neighbours is not the cause of either the stellar or \hi\ asymmetries observed in CIG\,85. \textcolor{black} {  Although CIG\,85 is a member of a loose group the large projected separation distances from other similar sized group members result in it having the isolation parameters  of a well isolated galaxy. On the other hand,  CIG\,85 has two blue stellar clumps which are reported as dwarfs galaxies in \cite{woods06}. \cite{woods06} suggest CIG\,85 is an interacting system \textcolor{black}{including} the two reported dwarfs,   020036+21480-b (02h 03m 18.8s +22d 02m 22.0s V = 2690 \km)  and 020036+21480-c (02h 03m 16.8s +22d 03m 01.0s V = 2618 \km) which are indicated with red circles  in  Figure \ref{sdss}. \textcolor{black}{These} two objects are projected within the  optical disk  of  CIG\,85 but are not  unambiguously separate galaxies.}

\begin{figure}
\centering
\includegraphics[scale=0.5]{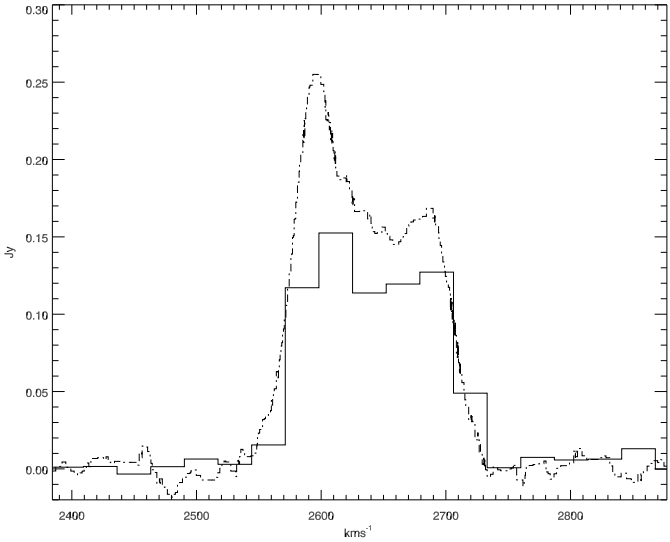}  
\caption{\hi\ spectra for CIG\,85: Single dish from \cite{tiff88}  (dashed line) and from GMRT (solid line).  }
\label{spect}
\end{figure}

 The  \lfir\ derived from the IRAS 60 $\mu$m  and 100 $\mu$m  fluxes falls within the normal range for an  isolated galaxy, see Figure 8 of \cite{lisen07}. \textcolor{black}{CIG\,85 \hi\  spectra have previously been obtained from the Westerbork Synthesis Radio Telescope (WRST) \citep[][49\prim\ angular resolution]{braun03} and  the Greenbank 91--m telescope   \citep[][10\prim\ FWHM beam]{tiff88}. The Tifft's spectrum is  shown with the  dashed line in Figure \ref{spect} and its  asymmetry suggests a lopsided distribution in the  \hi\  disk.} 
 Such a perturbed distribution of stars and \hi\ in an apparently isolated galaxy makes CIG\,85 an interesting candidate to study in detail.  Our \textcolor{black}{ \hi\ } mapping of the galaxy (see sections below) shows this galaxy to be  even more asymmetric \textcolor{black}{than \textcolor{black}{could} be inferred from the  \aflux\ } asymmetry parameter. This is because  the \textcolor{black}{ \aflux\ } ratio parameter misses a few cases where the shape of a real asymmetric profile \textcolor{black}{does not correspond to a difference in the total
areas of the approaching and receding sides \citep[][see for details]{espada11} .}


 We present in this paper results from the \textcolor{black}{GMRT} interferometric \hi\ and 20 cm radio continuum  observations of CIG\,85. \textcolor{black}{Sloan Digital Sky Survey, SDSS \citep{york00}}, and GALEX publicly available images have also been used in this paper. Section \ref{obs} sets out details of our observations with the results given in section \ref{results} and we discuss the possible reasons for the observed asymmetries in section \ref{discuss}. Our conclusions are set out in section \ref{conclusions}. J2000 coordinates are used through the paper.

\begin{table}
\centering
\begin{minipage}{190mm}
\caption{CIG\,85 parameters}
\label{table_a}
\begin{tabular}{lll}
\hline
\textbf{property}&\textbf{value} &\textbf{ reference } \\ 
V$_{optical}$&2640\,\km\ &\textcolor{black}{ \cite{vm05}}\\
RA&02h 03m 20.3s&Leon \& Verdes-\\
  & & Montenegro (2003) \\
DEC&+22d 02m 29.0s&\textcolor{black}{ \hspace{.1cm}" \hspace{1.5cm}" }\\
  & &  \\
Distance&35.9 Mpc&  \cite{fernandez12} \\
Spatial scale&$\sim$10.4 kpc/arcmin&\textcolor{black}{ \hspace{1cm}" \hspace{1.5cm}" }\\
Optical size&  2.0 x 2.0 arcmin&\textcolor{black}{ \hspace{1cm}" \hspace{1.5cm}" }\\
Morphology&Irregular&\textcolor{black}{ \hspace{1cm}" \hspace{1.5cm}" } \\
log($L_B$)& 9.47  \lsolar &\textcolor{black}{ \hspace{1cm}" \hspace{1.5cm}" } \\
log(L$_{FIR}$) & 8.86\lsolar&  AMIGA database\\
\hline
\end{tabular}
\end{minipage}
\end{table}



\begin{figure*}
\centering
\includegraphics[scale=0.6]{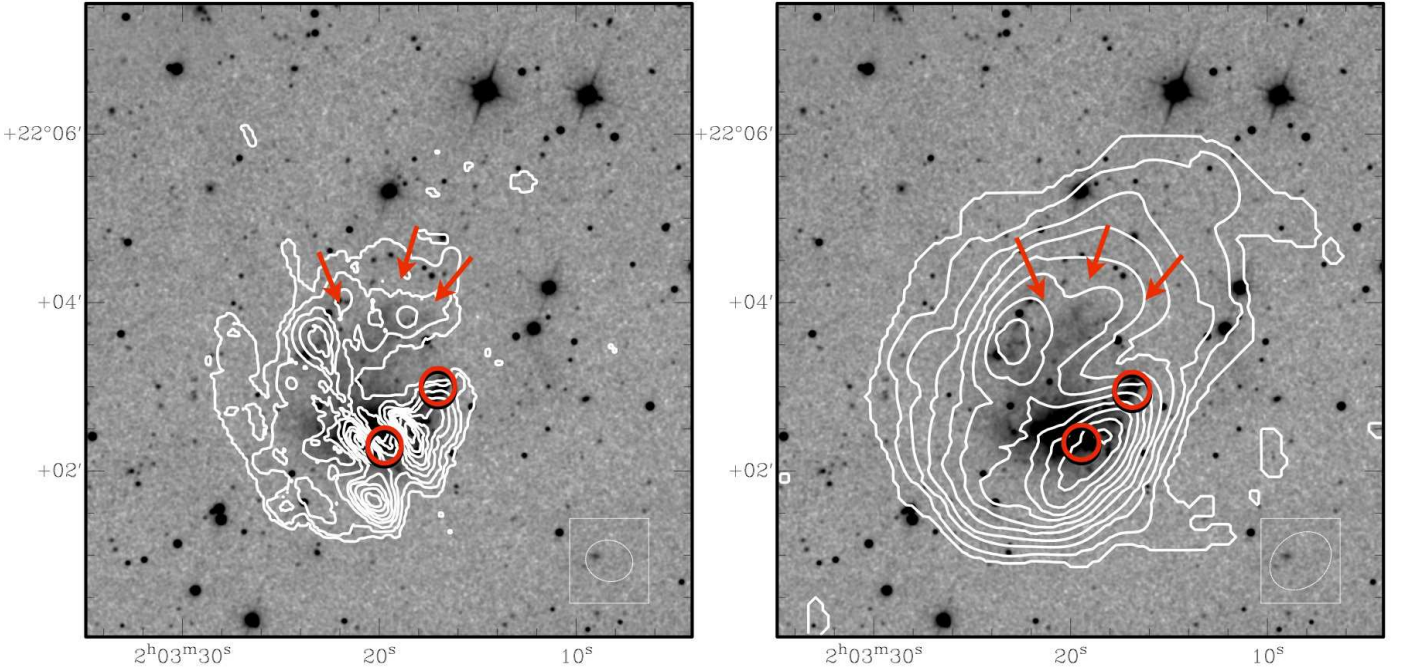}  
\caption{\textcolor{black}{Integrated  \hi\ contours on an SDSS \textit{r}--band image, where the three faint optical arm like features are
marked with red arrows and  020036+21480-b AND 020036+21480-c by circles. \textit{(Left) }High resolution integrated emission, (beam size = 24.3$^{\prime\prime}$ ′ x 21.0$^{\prime\prime}$ ) where the \hi\ column
density levels are 10$^{20}$ \textcolor{black}{\acm}  x (1.0, 1.9, 3.2, 5.3, 7.5, 9.6, 11.8, 13.9,
16.1)  \textit{(Right)} Low resolution integrated emission, (beam size = 47.1$^{\prime\prime}$  x 37.1$^{\prime\prime}$ ) where the
the \hi\ column
density levels are 10$^{20}$ \textcolor{black}{\acm}   x  (0.56, 1.2, 2.2, 3.4, 4.7, 6.0, 6.9, 8.2, 9.4, 11.0).
The beam is shown at the bottom right of each panel. }}
\label{mom0}
\end{figure*}

 \begin{figure}
\centering
\includegraphics[scale=0.5]{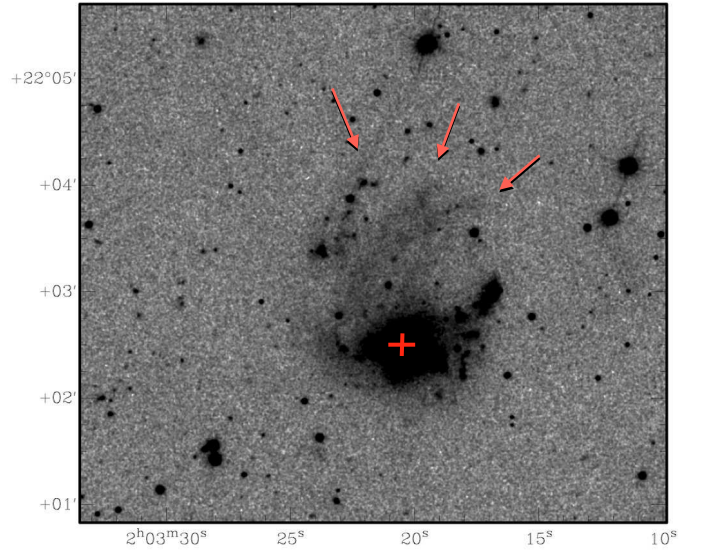}  
\caption{ SDSS \textit{r}--band image (25- 26 mag arcsec$^{-2}$) show\textcolor{black}{s} faint emission to the north of the main stellar concentrations. The  three faint plume or arm like features  are marked with red arrows and the cross indicates the optical centre.}
\label{fig2}
\end{figure}



\section{Observations}
\label{obs}
 21--cm \hi\ line and continuum emission from CIG\,85  was observed for 10 hours with the GMRT in November, 2009. 
The full width at half maximum (FWHM) of the GMRT primary beam at  1.420 GHz is $\sim$24\arcmin. 
The baseband bandwidth used  for  \hi\ line observations was 8 MHz giving a
velocity resolution of $\sim$13.7 \km\ in the velocity range of 1800 \km\ to 3500 \km. Observational parameters, including the
rms noise and beam sizes used to produce the integrated maps are   presented in Table \ref{table2}.


The  GMRT data were reduced using the Astronomical Image Processing System (\textsc{AIPS}) software package. 
Bad data due \textcolor{black}{to} malfunctioning antennas, antennas with abnormally low gain and/or  radio frequency interference (RFI)
were flagged.   The primary flux density calibrators used in the observations were 3C147 and 3C48, and  the phase calibrator was 0237+288 (see Table \ref{table2}).  The flux densities are on the scale
\textcolor{black}{of} \cite{baars77}, with flux density uncertainties of $\sim$5 per cent. The radio continuum images were  \textcolor{black}{generated from}  the line free channels. The \hi\ cubes were produced following continuum subtraction in the uv domain using the AIPS tasks \textsc{UVSUB} and \textsc{UVLIN}. 
The task  \textsc{IMAGR} was then used to obtain the  final cleaned  \hi\ cubes. From these cubes the integrated \hi\ and  \hi\  velocity field maps were extracted using the \textsc{AIPS} task  \textsc{MOMNT}. \textcolor{black}{To analyse
the \hi\ and radio continuum structures in CIG\,85 we produced
image cubes and maps of different resolutions by tapering \textcolor{black}{with different parameters}, and retained for this paper the ones with 
beam sizes of  24.3$^{\prime\prime}$  x 21.0$^{\prime\prime}$   and 47.1$^{\prime\prime}$  x 37.1$^{\prime\prime}$. It is to be noted here that the shortest spacing between GMRT antennas is$\sim$ 60 m which in L band
imposes an angular size limit of 6$^{\prime}$  to 7$^{\prime}$  for the observed sources.}
We note that some RFI could not be completely removed and remain\textcolor{black}{s} visible at low levels  in the maps.


\begin{table}
\centering
\begin{minipage}{130mm}
\caption{GMRT observation  details}
\label{table2}
\begin{tabular}{lll}
\hline

\hline

&\hi\ 21 cm line& radio continuum \\
&                           &\textcolor{black}{(1405.2026 MHz)}\\
Observation Date &13 November 2009& \\
Phase Calibrator & 0237+288 & 0237+288\\
Phase Calibrator & & \\
flux density&2.4 Jy \textcolor{black}{\km}&2.4 Jy \textcolor{black}{\km} \\
Integration time & 10 hrs& 10hrs \\
rms (per channel) & 1.15 mJy beam$^{-1}$& 0.33 mJy beam$^{-1}$\\
Beam (major axis) &24.3$^{\prime\prime}$& 47.2$^{\prime\prime}$ \\
Beam (minor axis) &21.0$^{\prime\prime}$ &37.1$^{\prime\prime}$ \\
PA &15.8\degree& -39.4\degree \\

\hline
\end{tabular}
\end{minipage}
\end{table}



\section{Observational Results}
\label{results}

\subsection {\hi\ content and distribution}
\label{hi_content}

Figure \ref{mom0}  shows the GMRT integrated \hi\ maps ($\sim$ 25$^{\prime\prime}$  and ∼ 45$^{\prime\prime}$  resolution)
overlayed on an \textcolor{black}{SDSS} \textit{r}--band image. The high resolution \hi\ map (\textit{left panel}) shows the 
\hi\ disk to
be highly disturbed with a clumpy distribution\textcolor{black}{, with the large majority of the projected disk area having a   }  column density $<$ 
10$^{21}$ \acm. The high density \hi\ (with maximum column densities of
∼ 1.6 x 10$^{21}$ \acm )~ is principally located in two clumps
whose maxima are located ∼ 50$^{\prime\prime}$  (8.6 kpc) south and ∼ 30$^{\prime\prime}$  (5.2
kpc) west of the optical centre. The SDSS \textit{r}--band image (Figure \ref{fig2}) shows three prominent optical extensions (marked with red
arrows) north of the optical centre. The \hi\ disk, measured at
a column density of $\sim$ 2 x  10$^{20}$  \acm  extends well beyond the
optical extent of the galaxy, with the high density clump in the
northern half of the disk coinciding with the \textcolor{black}{easternmost} of the
optical extensions.
The lower resolution \hi\ map (\textit{right panel} Figure \ref{mom0}) provides a complementary picture
showing more obviously the large extent of the atomic gas in the north--west \textcolor{black}{which continues} well beyond the three optical extensions. \textcolor{black}{The background image in this figure has  been overexposed for clarity.} However the  \hi\ in the far NW seems to be a continuation of the optical
features. Deep optical observations of CIG\,85
may reveal the presence of extended faint stellar structures in this north-western area. \textcolor{black}{alternatively the absence of such structures could be used to argue for cold gas accretion, as was done in  NGC\,3367 where its asymmetric morphology and nuclear activity was attributed to cold gas accretion \citep{toledo2011}.}


 The integrated \hi\ flux density obtained from \textcolor{black}{Greenbank 91--m single dish observations of  \cite{tiff88} }is 27.3 Jy \km, implying an \hi\ mass for the galaxy of 8.3 $\times$ 10$^{9}$ M$_\odot$. Comparing this to the \hi\ mass observed by the GMRT (5.8  $\times$ 10$^{9}$ M$_\odot$) indicates the interferometer is only recovering $\sim$ 70\% of \hi\ flux and implies that CIG\,85 contains a large mass of diffuse \hi. We can identify where this flux difference comes from by looking at Figure \ref{spect}. There the \hi\ spectrum from our GMRT observations (solid line) is compared to the single dish spectrum (dashed line) from \cite{tiff88}. The single dish spectrum displays a double horn profile, with a larger \hi\ flux in the low velocity horn ($\sim$ 2580 \km) compared to the high velocity horn ($\sim$ 2700\km). Our lower velocity resolution (26 \km) GMRT interferometric spectrum (solid line in Figure \ref{spect}), while having a similar velocity and velocity width, is missing flux across the range of detected velocities. However the loss is greatest in the low velocity horn and our \hi\ velocity field (Figure \ref{velfield}) shows the low velocity \hi\ is located in the NW, where the \hi\ disk is most disturbed. For this reason in all further calculations requiring the \hi\ mass, we use the single dish value. Taking the major diameter (Table \ref{table_a}) as $\sim$ 2.0\textcolor{black}{$^{\prime}$} gives a log${{\frac{M_{H_{I}}}{D_{l}^{2}}}}$ value of 7.16. A comparison with the \hi\ surface densities of  isolated  galaxies of similar morphological type \citep{hayn84} shows that  CIG\,85 has normal \hi\ content. The AMIGA sample is more isolated than the  \cite{hayn84} sample so we compared the single dish \hi\ mass  with our AMIGA sample as well. The comparison with the AMIGA sample confirms the \hi\ content of CIG\,85 is  similar to \textcolor{black}{{ that of} an isolated galaxy of its morphological type and size } (Espada et al. in preparation).




The AMIGA isolation parameters are based on 2D optical data and  may not include optically faint gas rich companions. Therefore, the \hi\ cube was searched over the entire  GMRT primary beam, (the L band FWHM being $\sim$ 24$^{\prime}$ which is $\sim$ 250 kpc at the distance of CIG\,85),  and $\sim$ 1400 \km\ velocity range,  for previously undetected companions.  Although the signal to noise ratio (SNR) drops considerably beyond the primary beam FWHM, an area twice the FWHM diameter of ($\sim$0.5 Mpc)  was also searched but no \hi\  companions were detected in either search. The \hi\ mass detection limit inside the FWHM of the GMRT primary beam, assuming a  3 channel line width and 3 $\sigma$ SNR, is $\sim$ 8.5 $\times$10$^{7}$ \msolar. \textcolor{black}{It is interestesting to note here that  \cite{braun03} spectrum shows strong \hi\ emission at velocities of $\sim$ 2800 \km\  which are not present in either the \cite{tiff88} or our own GMRT spectra. The effective beam size of these observations was $\sim$ 49$^{\prime}$. However there were baseline problems with the spectrum in this velocity  range (Braun private communication) and the position from which the spectrum was extracted was not stated. Therefore, though a striking feature near 2800 \km\  is present in this spectrum, no conclusion can be drawn from it.}

\subsection {\hi\ velocity field and rotation}
\label{velo}

\begin{figure}
\centering
\includegraphics[scale=0.55]{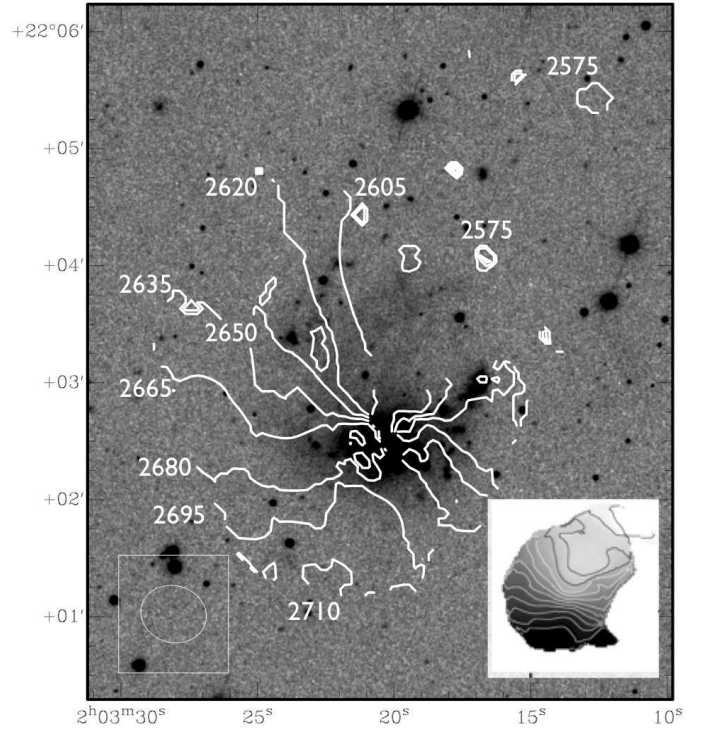}  
\caption{\hi\ velocity field contours \textcolor{black}{derived from the high resolution cube on SDSS \textit{r-}-band band image (main panel).} The velocity contours are plotted in  15 \km interval. The spatial resolution is 24.3 $^{\prime\prime}$ $\times$ 21.0 $^{\prime\prime}$ with the beam size  indicated by the ellipse. \textcolor{black}{(Inset panel:)  Low resolution \textcolor{black}{(47.1 $^{\prime\prime}$ $\times$ 37.1 $^{\prime\prime}$),} systemic velocity subtracted, \hi\ velocity field  for CIG\,85. The systemic radial velocity is 2640 \km. From white to black, the contours (in \km\ ) are 2570, 2580, 2590, 2600, 2610, 2620, 2630, 2650, 2660, 2670, 2680, 2690, 2700, 2710,  2720. The velocity resolution of the cube is $\sim$ 26 \km.}}
\label{velfield}
\end{figure}

In Figure \ref{channel_maps} we display the channel maps corresponding to the \hi\ emission at the indicated heliocentric velocities with a channel width of $\sim$26  \km\  and an rms noise of 0.012 Jy \textcolor{black}{beam$^{-1}$}, implying a detection threshold level of 6.7 $\times$ 10$^{20}$ \textcolor{black}{\acm.} The \hi\ contours have been overlayed on the \hi\ integrated map displayed in  grey scale. \textcolor{black}{The channel maps are rather regular in velocity while clumpy in distribution,} the best behaved being the channels closest to the systemic velocity. The redshifted channels show most of the \hi\ mass is located in the eastern part, also reaching higher column densities there. In the velocity range from 2646.8 \km\ to 2687.1 \km\, the faintest eastern emission coincides with part of the eastern optical \textcolor{black}{"plume"} closer to the center of the galaxy (Figure \ref{fig2}). On the other hand, the blueshifted \hi\ is significantly more prominent in the western side where it seems to cover the outer (larger radii) parts of the three optical plumes. The velocity field\textcolor{black}{, derived from the \hi\ channel maps,} is presented in Figure \ref{velfield}. \textcolor{black}{The main panel of Figure \ref{velfield} shows the \hi\  velocity field derived from the high angular resolution cube, while the inset in the 
lower right corresponds to the low resolution one,  highlighting the
            kinematics of the more extended atomic gas.} Like the channel maps, the velocity field also shows a larger regularity in the central parts at \textcolor{black}{velocities closest  to the }systemic one. There it is consistent with a rotating disk. The northern half shows \textcolor{black}{fewer} isocontours due to low signal to noise \hi\ detection\textcolor{black}{s} in several places in the NW of the disk. Hints of a warped \hi\ disk are seen in the iso–velocity contours in the SE and southern edge and also in the channel maps (Figure \ref{channel_maps}). NW of the stellar plumes the \hi\ is particularly disturbed with discrete clumps of gas extending as far as $\sim$ 02h 03m 14.0s +22d 05m 30.0s with no apparent optical counterpart in the SDSS images, but maintaining a systematic gradient in velocity with radial distance (2606 \km\ to 2575 \km). As noted in section 3.1, most of the diffuse emission ($\sim$ 2 $\times$ 10$^{9}$ M$_\odot$)  is expected to be located in this region. The channel maps (Figure \ref{channel_maps}) clearly show the \hi\ disk to be stretched in the E and NW. \textcolor{black}{The kinematic centre of the galaxy is at 02h 03m 20.3s, +22d 02m 39.8s, i.e. within the uncertainties at the optical centre.} 

In Figure \ref{pvdiag} we present a position-velocity cut along a position angle of $\sim$146\degree . The direction has been selected as a compromise to include the northwestern  \hi\ blob,  being as close to the major kinematical axis of the galaxy as possible, and avoiding the  RFI effect at the galaxy centre. The zero point in the figure corresponds to  02h 03m 21.2s  and +22d 02m 47.5s, while the optical centre is 02h 03m 20.3s +22d 02m 29.0s and the kinematic centre from the moment 1 map is estimated to be at 02h 03m 20.3s, +22d 02m 39.8s.  
Surprisingly,  the approaching side of the position-velocity diagram is consistent with regular rotation, although this part of the galaxy appears to be more  perturbed in the velocity field. The exception is the northwestern clump,  which shows higher velocities than expected, suggesting it to be stripped gas. On the other hand  the rotation curve in the receding side increases  in  a more irregular way.

\textcolor{black}{From the outline of the \hi\ distribution we estimate the inclination of the galaxy is $\sim$ 45\degree\ and from the velocity width V = $\frac{V_{obs}}{sin(i)}$}, we estimate the V$rot$ =100 \km implying a dynamical mass of $\sim$ 4.6 $\times$10$^{10}$ M$_{\odot}$. Making the further assumption that the mass of molecular gas is $\simeq$  M(\hi), the  baryon fraction of CIG\,85 would be $\sim$ 0.35, implying a normal dark matter content similar to that found in spiral galaxies.




\subsection {20 cm radio continuum from CIG\,85}
\label{20cm}

20 cm radio continuum was  detected  with a total  flux density of 2.2 mJy.  The lower panel of Figure \ref{uvha} shows the low resolution  radio continuum emission from CIG\,85, overlayed on SDSS \textit{ r}--band image. The upper and middle panels of Figure \ref{uvha} respectively show the composite images of GALEX NUV contours with $\sim$ 5.3 arcsec resolution and H$\alpha$ contours from \cite{ james04} with resolution $\sim$ 1 arcsec, both overlayed on SDSS \textcolor{black}{ \textit{ r}--band} image. The radio  continuum emission shows a significant overlap with the other recent SF  tracers, \halpha\ and UV (Figure \ref{uvha}). The star formation rate (SFR), estimated from the 20 cm data, is  0.18 \msolar\ yr$^{-1}$  \citep{yun01} and that  from the 60 $\mu$m  data is $\sim$ 0.5 M$\odot$ yr$^{-1}$  \citep{2002AJ....124..862H}. Assuming that the entire radio continuum emission is non--thermal, the SN rate is calculated to be  0.003 yr$^{-1}$  \citep{1990ApJ...357...97C}. The FIR--radio  relation \citep{yun01} parameters for CIG\,85 are, \textcolor{black}{log L($_{1.4 GHz}$) = 20.5  and log( L$_{60 \mu m}$) } = 8.6,    confirming that the galaxy  lies on the FIR--radio correlation  \citep{2008A&A...486...73S}.


\section{Discussion}
\label{discuss}

In this section we first discuss the environment, minor companions, stellar  characteristics and morphology of CIG\,85, to understand the properties of this system. Then we compile  all the stellar and H{\sc i} observational evidence and reach  a conclusion about which of the processes listed below can explain most of the observations and therefore is  the likely cause of the current  \hi\ and stellar  asymmetries of CIG\,85. As an isolated galaxy, CIG\,85, could be affected by any one or a combination of the following processes, which can give rise to its asymmetry.
\begin{itemize}
\item[1]
Accretion of a large mass of cold gas or \textcolor{black}{High Velocity Clouds (HVCs)}.
\item[2]
Internal perturbations within the disk driven by a bar.
\item[\textcolor{black}{3}]
Intense star  formation in the disk.
\item[\textcolor{black}{4}]
Recent accretion of a \textcolor{black}{satellite companion} galaxy.
 \end{itemize}

\begin{figure*}
\centering
\includegraphics[scale=0.85]{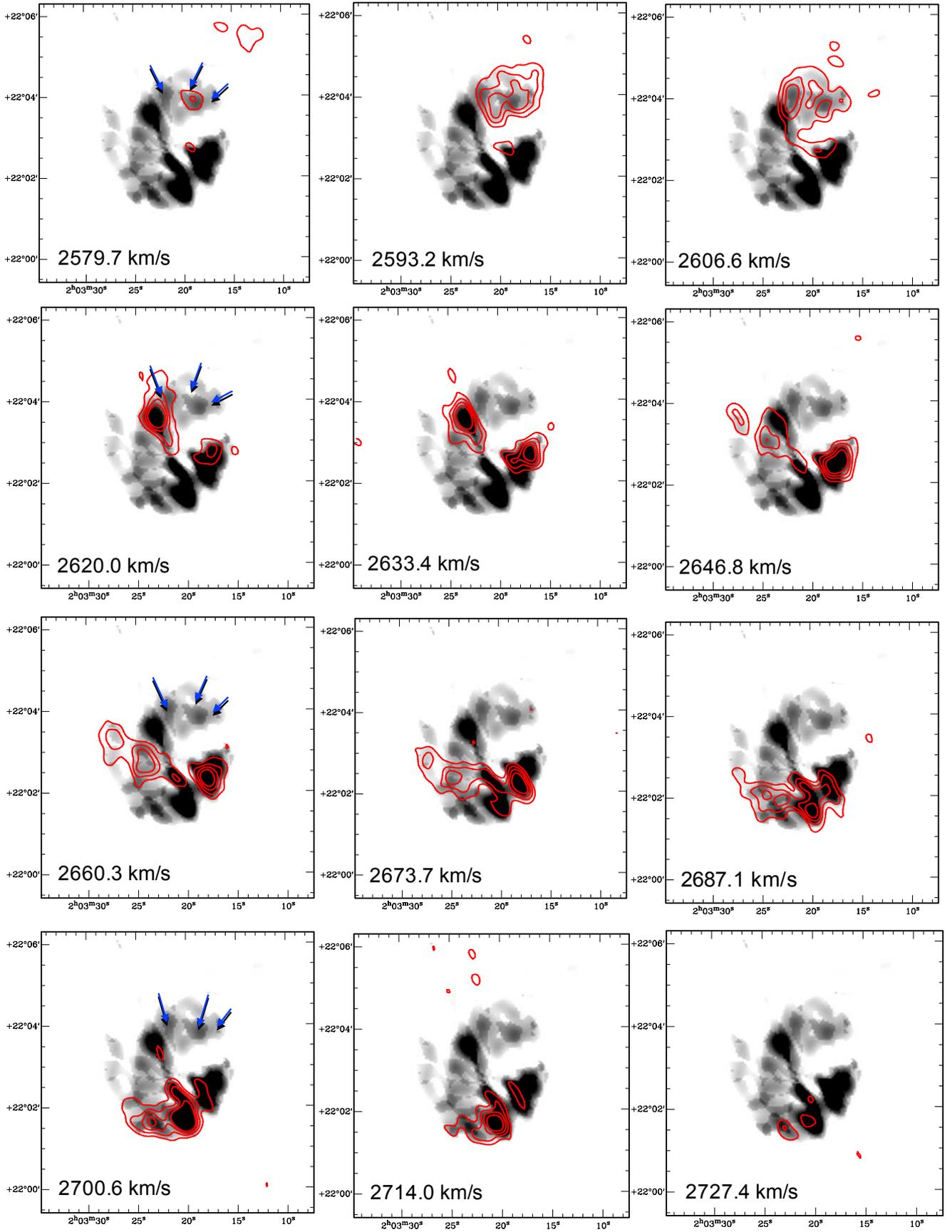}  
\caption{Channel images (red contours) at a spatial  resolution of 24.3 $^{\prime\prime}$ $\times$ 21.0 $^{\prime\prime}$ overlayed on grey scale integrated H{\sc i} map. The velocity in \km\ is shown in the bottom left corner of each frame. Contour levels are at 3,5,7 and 9 $\sigma$ where  $\sigma$  corresponds to  1.4 K.  Leftmost boxes of each row have three blue arrows on the galaxy to point the positions of the optical arm like features.}
\label{channel_maps}
\end{figure*}

\begin{figure*}
\centering
\includegraphics[scale=0.65,angle=-90]{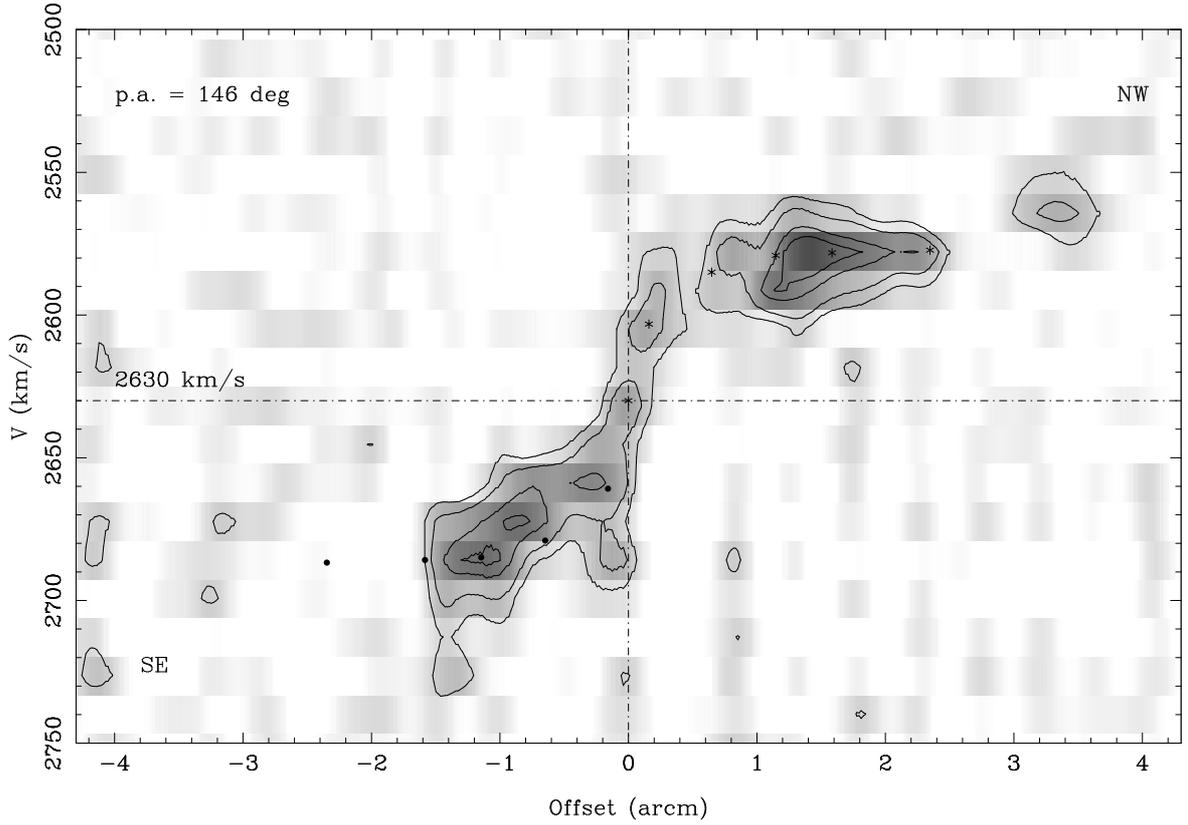}  
\caption{Position-velocity cut along a position angle of $\sim$146\degree . The zero point in the figure corresponds to  02h 03m 21.276s  and 22 02 47.59. The  vertical line corresponds to this zero position, and 
the horizontal line to the velocity that maximises symmetry between the receding and approaching sides of the curve (in this particular cut, 2630 km/s). The stars trace the approaching side and the dots the receding side, in order to illustrate the differences between each side.  }
\label{pvdiag}
\end{figure*}

\subsection{Minor companions }
\label{companions}
 

\textcolor{black}{As has been mentioned in the Introduction, 020036+21480-b and 020036+21480-c, indicated with red circles  in  Figure \ref{sdss}, have been reported as dwarf companions of  CIG\,85 by \cite{woods06}. However, the entities are projected within the disk of CIG\, 85, so they cannot be unambiguously categorised as separate galaxies.} The AMIGA isolation parameters given in the Introduction  are derived taking into account only similar sized galaxies, i.e. are designed to identify presence or absence of companions capable of \textcolor{black}{producing} a  major interaction with the galaxy. If the tidal force parameter Q$_{kar}$ for CIG\,85 is estimated taking into account  all \textcolor{black}{of} its close neighbours, without any restriction on their size,  the effect of the two dwarfs become\textcolor{black}{s} significant. The value of Q$_{kar}$ becomes $>$ 0  (\textcolor{black}{well} above the AMIGA cut--off value of $\le$ -2.0) indicating that the external tidal forces, mainly created by the very nearby companions,  would be higher, and clearly dominate over the internal binding forces. So, provided the \cite{woods06}  dwarfs are indeed entities separate from CIG\,85, minor companions could be the explanation for the asymmetries observed in CIG\,85.  

\begin{figure}
\centering
\includegraphics[scale=0.70]{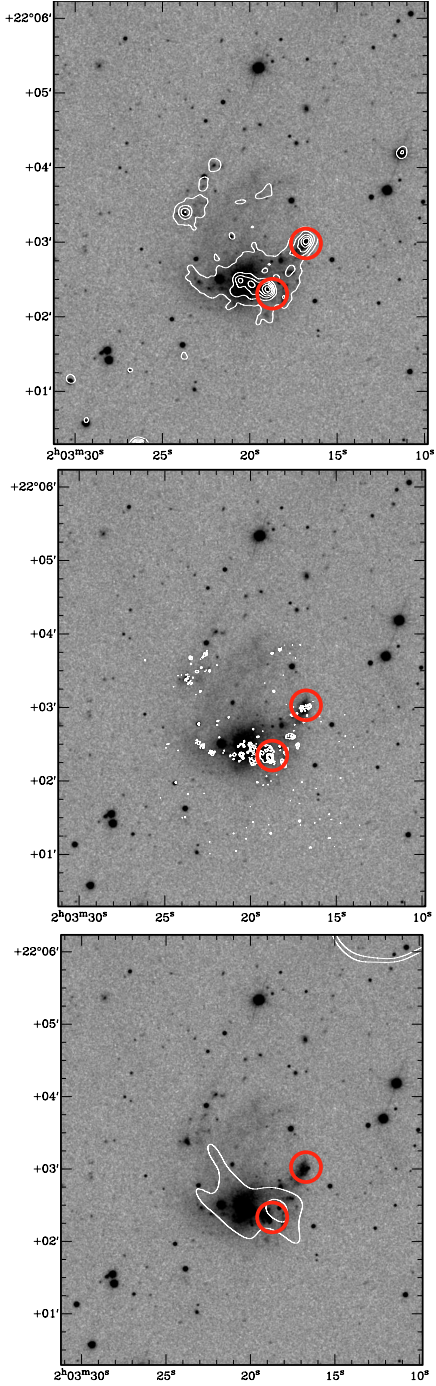}  
\caption{Composite image of GALEX NUV contours with $\sim$ 5.3 arcsec resolution (top panel), H$\alpha$ contours from \cite{ james04} with resolution $\sim$ 1 arcsec (middle panel) and 20 cm radio continuum contours with resolution $\sim$ 45 arcsec  (lower panel), each overlaid on an SDSS-r band image. The circles indicate  020036+21480-b and 020036+21480-c }
\label{uvha}
\end{figure}

UV (\textit{GALEX})  images (Figure \ref{uvha}) show the brightest SF regions within the CIG\,85 disk are co--spatial with  the  positions of the dwarfs and \textcolor{black}{ the SDSS spectrum for 020036+21480-b has the characteristics of a low abundance entity. Estimated oxygen abundance from the SDSS spectrum for this object is low, about 0.2 solar abundance, typical of dwarf galaxies or the Small
Magellanic Cloud.} However, we  only have indirect evidence suggesting the  \textcolor{black}{two dwarf galaxies reported by \cite{woods06}}  are individual galaxies or galaxy fragments rather than  regions of high SF within CIG\,85 disk, e.g. the presence of SDSS \textit{z}--band emission at positions of  the dwarfs is consistent with separate galaxies or fragments of satellite galaxy, rather than merely young CIG\,85 \hii\ regions. \textcolor{black}{So these regions being part of the CIG\,85 cannot be ruled out as well.}

\begin{figure}
\centering
\includegraphics[scale=0.6]{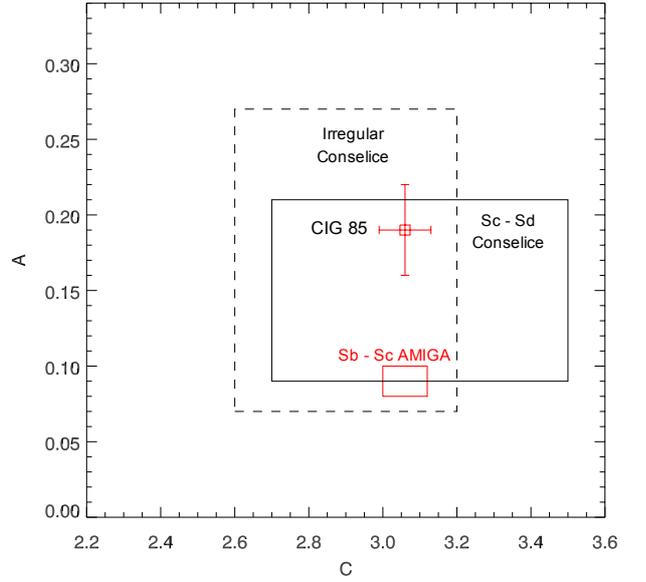}  
\caption{CIG\,85 \textit{r}--band C (concentration) and A (asymmetry)  parameters  with error bars-- red point. The range of  A and C CAS parameters for Sb--Sc spirals (sample size $\sim$100) from  the AMIGA sample \citep{durbala08} are indicated with  the red rectangle.  The range of the same  A and C parameters from \cite{consel03} is displayed for irregular and Sc--Sd morphologies respectively as a dashed black rectangle  and a solid black rectangle.}
\label{cas}
\end{figure}

\subsection{Stellar  properties}
\label{stellar}
The stellar mass estimated from the \LB, following the method from  \cite{Bell}, is 2.4$\times$10$^{9}$ M$_{\odot}$.  This compares well with \textcolor{black}{the} stellar masses derived from the Spitzer IRAC 3.6 $\mu$m and 4.5 $\mu$m bands which give  1.7$\times$10$^{9}$ M$_{\odot}$ and 1.3$\times$10$^{9}$ M$_{\odot}$ respectively. Overall, the stellar mass distribution  is  strongly asymmetric  (based on SDSS images), with most of the mass in the southern \textcolor{black} {side} of the galaxy, while the northern part of the disk is  more diffuse  (Figures  \ref{fig2} \& \ref{mom0}).  The SDSS \textit{u, g, r, i } and \textit{z} band images indicate a clear segregation between the distribution of the  old and young stellar populations (Figure \ref{sdss}).The old population is most prominent \textcolor{black}{in the central region} and at lower densities \textcolor{black}{ in faint plume--like features running north from the main concentration of stars, which could be either spiral arms or debris from an interaction, (Figure \ref{fig2}).} In contrast the   young blue stars are predominantly found  in an arc which initially follows the \textcolor{black}{easternmost} plume (Figure \ref{fig2}) and  continues \textcolor{black}{through} the optical centre to the position of the smaller  \textcolor{black}{ of the  two dwarfs  reported by  \cite{woods06}}, 020036+21480-c (Figure \ref{sdss}).  As expected\textcolor{black}{,} the  UV (\textit{GALEX}), \halpha, radio continuum emission essentially  follow the distribution \textcolor{black}{of} the young blue stars (Figure \ref{uvha}). \textcolor{black}{Both the red stellar plume and the arc of young blue stars have somewhat similar distributions to  NGC\,922, a collisional ring galaxy, where an off-centre high velocity collision with a companion is thought to be responsible for the C -- shaped  \halpha\ morphology and the red extended optical plumes \citep{2006MNRAS.370.1607W}. However the  \halpha\ emission in CIG\,85 is dominated by the two clumps projected near the centre in contrast to  NGC\,922, where the emission is almost exclusively in the ring.}
 
\textcolor{black}{To place the optical morphology of CIG\,85 in perspective relative to isolated AMIGA and field galaxies of similar morphological types,} we  estimated the asymmetry (A) and concentration (C) parameters in the CAS  system for CIG\,85. It included the  ``minimisation of A''  step required for galaxies displaying an irregular morphology, from the SDSS \textit{r}--band  image using the method from \cite{consel03}.  The estimated asymmetry parameter,   A = 0.19 $\pm$ 0.03, is plotted in Figure \ref{cas}. In order to gain a perspective on the CIG\,85 A and C parameters, Figure \ref{cas} also shows the 1 $\sigma$ range of these parameters for AMIGA Sb--Sc spirals,  \citep[A = 0.09 $\pm$ 0.01][]{durbala08} -- red rectangle. The figure also shows the 1 $\sigma$  parameter ranges for  Sc--Sd spirals (solid black rectangle) and irregulars (dashed black rectangle) for the non-environmentally selected \textcolor{black}{sample} from \cite{consel03}. While the  CIG\,85 \textcolor{black}{CAS -- }C parameter is very close to \textcolor{black}{that of the} AMIGA spirals, the A parameter is clearly significantly larger than \textcolor{black}{for}  the AMIGA  spirals. The CIG\,85 \textcolor{black}{CAS -- }A parameter \textcolor{black}{also falls within} the upper half the Conselice A parameter space for both  irregular and spiral galaxies \textcolor{black}{and the Conselice sample is} known to include galaxies impacted by interactions.

\textcolor{black}{ The radial optical surface brightness profiles of a sample of late--type spirals studied by  \cite{pohlen06} generally show an initial exponential decline associated with the bulge followed by a shallower decline in the disk. For Hubble types later than Sd the Pohlen profiles show a \textcolor{black}{smaller} bulge contribution leaving the profiles dominated by the more shallowly declining disk component.   The surface brightness profiles  for CIG\,85 derived from the SDSS \textit{g,r,i}--band images (after removal of foreground stars) all \textcolor{black}{lack the bulge emission profile component  typical of}  spirals with morphologies in the range Sb to Sc}. \textcolor{black}{However Figure \ref{cas} shows }the CAS -- C parameter \textcolor{black}{is similar to that of } AMIGA \textcolor{black}{spirals with morphologies in the range Sb to Sc}. \textcolor{black}{The most likely explanation for the apparent inconsistency between the CIG\,85 CAS -- C parameter (consistent with a late--type spiral galaxy) and the light profile more consistent with  an irregular galaxy is the presence, near the optical centre,  of  the  two  stellar clumps reported as dwarfs galaxies in \cite{woods06} which would artificially increase the concentration parameter.}

\subsection{Morphology of CIG\,85: irregular or spiral?}
\label{morph}
Besides asymmetry, another interesting aspect of this system is its  morphological type. A detailed study of its properties indicate that CIG\,85 may not be a simple irregular galaxy. Several of its properties mimic a spiral,  suggesting it could be a galaxy whose morphology is currently undergoing a transformation. With an optical diameter of \textcolor{black}{$\sim$} 21 kpc CIG\,85  has a significantly  greater size than the median for UGC Sm--Im  irregular galaxies where D${25}$ = 17kpc  \citep{roberts94} and in general irregular galaxies are smaller in size than  spirals  \citep{pilyu04,roberts94}.  Other properties  which are more consistent with CIG\,85 being an Sc -- Sd spiral \textcolor{black}{rather} than an irregular include,  the log(L$_{B}$) = 9.15 [\lsolar]\textcolor{black}{,}  log(L$_{FIR}$) = 8.75  [\lsolar]  \citep[][UGC  sample]{roberts94} \textcolor{black}{and its  CAS -- C parameter.} The \hi\ mass, 7.9$\times$10$^{9}$ M$_{\odot}$, of CIG\,85  is also more typical of a spiral  than an irregular galaxy  \citep{roberts94}.  \cite{pilyu04}, while studying a sample of spirals and irregulars, found a clear difference between the  metallicity, rotational velocity and blue band magnitudes of irregulars and spirals. Comparing CIG\,85's M$_B$ which is \textcolor{black}{-}17.98 and  V$_{rot}$ $\sim$ 100 \km\  to the sample of spirals and irregulars in  \cite{pilyu04} we find that CIG\,85 has properties of a galaxy  lying  on, or just on the spiral side, of the irregular / spiral transition (See figs. 9, 10 and 11 of \cite{pilyu04}). These intermediate properties could be evidence that CIG\,85 is undergoing a transformation from a spiral to an irregular or vice versa, driven by one or more of the mechanisms proposed above.

\subsection{Star formation rate and \hi\ correlation}
\label{sfr}

The current global SFR estimated using the 60$\mu m$ data  is  $\sim$0.5 M$_{\odot}$ yr$^{-1}$ \citep{2002AJ....124..862H} \textcolor{black}{and SFR(\halpha) $\sim$0.78 M$_{\odot}$ yr$^{-1}$ \citep{james04}}. Assuming the galaxy has been forming stars over 13 Gyr,  the mean SFR would be 0.13 M$_{\odot}$ yr$^{-1}$   indicating the current  SFR is above the long term mean level. This and the fact that the  regions of strongest recent SF(UV)  are co-spatial with the location of the  dwarfs \textcolor{black}{reported by \cite{woods06}}, support the interaction induced star formation scenario. As was mentioned in section 4.2, the stellar mass of CIG\,85 is  even lower than its H{\sc i} mass. Also the major SF sites are close to the optical
and \hi\ kinematic centre of the galaxy and are hardly found in
other parts of the disk, except for the eastern optical plume
marked with an arrow in Figure \ref{fig2}.
These facts seem to  rule out the possibility that SF activity alone is responsible for the observed large scale \hi\ and stellar asymmetries. The current state of star formation in CIG\,85 is most probably an after-effect of the asymmetries rather than the cause of \textcolor{black}{them}.  
 \textcolor{black}{As was noted previously, there is no marked correlation of high \hi\ column
density regions with the star forming regions. There are three
clumps of \hi\ in the disk as seen in the high resolution map (Figure \ref{mom0} \textit{left panel})
having column densities $\geq$ 1.0 $\times$ 10$^{21}$  \acm. The two main clumps with column densities significantly
$>$ 1.0 $\times$ 10$^{21}$  \acm, are located towards the south. One of
them correlates well with the star forming possible dwarf galaxy
020036+21480-b and the other clump is well towards the south,
away from active star forming zones. The third clump having a
maximum column density $\sim$ 10$^{21}$  \acm is in the NW part
of the disk, where a blue SF region 
is seen in Figure \ref{sdss}, although with some shift with respect to its peak. However the
 area looks patchy in the SDSS images, suggesting it is suffering
from dust extinction which could explain this shift.}


\subsection{What could have happened to  CIG\,85, leading to its current morphology }
\label{sum}

\textcolor{black}{After carefully considering all the observational results presented above, we conclude that \textcolor{black}{accretion of a minor companion } most readily explains the current \hi\ and stellar asymmetries observed in CIG\,85. Under this scenario, originally CIG\,85 was a gas rich spiral with a low SFR. \textcolor{black}{Over the } course of time the orbit of a smaller companion has decayed to the point where it has deposited \textcolor{black}{tidal} debris in and near the parent galaxy \textcolor{black}{(e.g. the plumes seen in the \textit{r} -- band image - Figure \ref{fig2}). }  } \textcolor{black}{The  interaction with the satellite } has resulted in the current perturbed and lopsided \hi\ and optical distributions. In this scenario the arc of blue stars from the east to centre would trace shock driven star formation  arising from accretion of fragments of the companion, which is in a late stage of being subsumed into CIG 85.  To reach this conclusion we have considered the following questions:

\begin{itemize}

\item
Can the mechanism explain both the  systematic large scale differences between diffuse northern and denser southern distributions, in both the  \hi\ and stellar components,  while simultaneously accounting for the lack of correlation between the older stellar, younger stellar and \hi\ high density regions on smaller scales?
\item
Are the two brightest SF regions  either dwarfs/interaction remnants or SF regions within the CIG\,85 disk?
\item
What is the interpretation of the rotation pattern in the \hi\ disk?
\item
Can the scenario explain  the presence of low metallicity, but not pristine, gas  in the strongest SF region?   

\end{itemize}

The arguments supporting  the scenario described above are summarised next:
\begin{enumerate}
\item
The CAS -- A  parameter which quantifies the asymmetry, is very \textcolor{black}{large} compared to that of the AMIGA spirals and in the upper quartile of the CAS -- A  parameter space \textcolor{black}{of} the  \cite{consel03} spiral and irregular galaxy sample\textcolor{black}{s}, which is indicative of a strong recent perturbation.

\item
\textcolor{black}{ The } detailed comparison with \cite{roberts94} and \cite{pilyu04} \textcolor{black}{ suggests} \textcolor{black}{that CIG\,85 resembles in a number of ways} a gas rich spiral undergoing a morphological type transition rather than an irregular.

\item
The SF history of CIG\,85 \textcolor{black}{suggests} a modest SFR of 0.13 M$_\odot$ yr$^{-1}$  in the past
compared to a higher current SFR of 0.5 M$_\odot$ yr$^{-1}$.  \textcolor{black}{Enhanced SF is a known consequence of interactions.}

\item
 \textcolor{black}{The extended \hi\ to the \textcolor{black}{NE} shows continuity with the velocity gradient of the galaxy, as well as the eastern emission  visible in the \hi\ channel maps (Figure \ref{channel_maps} ) from 2646.8 \km\ to 2673.7 \km\ consistent with  these extensions being perturbed parts of the main body of the galaxy.}


 \item
\textcolor{black}{The metalicity of the gas in the  020036+21480-b region   is $\sim$ 0.2 solar, consistent with that expected in objects such as SMC or a dwarf galaxy. Given the massive  \hi\ mass of CIG\,85 compared to its optical mass and its low SFR history, it is not suprising that CIG\,85 will be metal poor. Additionally, this is also consistent with the idea that 020036+21480-b region is part of a small metal poor companion galaxy whose accretion into  CIG\,85 is responsible for the current asymmetry.}

 \item 
The persistence of the large scale rotation pattern in the \hi\ disk and the coincidence of the projected \hi\ kinematic and optical centres of the galaxy, indicates the mass of the  companion was insufficient to destroy the overall \hi\ rotation pattern but was sufficient to significantly perturb the \hi\ morphology.

 
\item
The coincidence of the brightest star forming regions with  020036+21480-b  and 020036+21480-c reported by \cite{woods06} as dwarf galaxies, is both consistent with accretion of a satellite and interaction induced star formation and 
also provides an explanation for the morphologies of  	red (pre-existing small bulge)  and 	blue  (blue stellar arc) stellar components.

\end{enumerate}

\textcolor{black}{The \aflux\ \textcolor{black}{asymmetry parameter} of the AMIGA galaxies for which we have carried out \textcolor{black}{spatially}  resolved \hi\ studies \textcolor{black}{ as part of the AMIGA project, i.e. CIG\,85, CIG\,96 \citep{espada05,espada11,2011ApJ...736...20E} and CIG\,292 \citep{portas}, } are  1.27, 1.16 and 1.23 respectively. Two  features common to  all three galaxies are the  indication that the \hi\ asymmetries were generated \textcolor{black}{relatively} recently ($\sim$ 10$^8$ yr ago) and that they were not produced by interactions with major companions. Additionally in all three cases the  \hi\ velocity fields show overall regular rotation, but with \textcolor{black}{indications that}  \hi\ disks contain a warp and  the signatures of  \hi\  asymmetry are  more apparent in the 2D distributions than in the velocity fields.  While the balance of evidence favours accretion of a minor companion as the cause of the \hi\ asymmetries in CIG\,85 and  interaction with a minor companion in the case of CIG\,96, it was concluded that the most likely cause of the asymmetry in CIG\,292 was an internally produced perturbation. }

\section{Conclusions}
\label{conclusions}
Thus evaluating all the observational evidence, we come to a conclusion that CIG\,85 is most likely a case of disturbed spiral galaxy which now appears to have the morphology of an irregular galaxy. Though it is currently isolated from major companions, the observational evidence is consistent with \hi\  asymmetries, highly disturbed optical disk and recent increase in star formation having been caused by a minor  merger, remnants of which are  now projected in front of the optical disk. If this is correct, the companion will be fully  accreted by CIG\,85 in the near future.\\
\\

{\small    {\bf Acknowledgments} We thank the staff of the GMRT who have made these observations
  possible. GMRT is run by the National Centre for Radio Astrophysics
  of the Tata Institute of Fundamental Research. We thank the reviewer for his/her useful comments and suggestions. This work has been supported by Grant AYA2008-06181-C02 and AYA2011-30491-C02-01,
cofinanced by MICINN and FEDER funds, and the Junta de Andalucia (Spain) grants P08-FQM-4205 and TIC-114.
The \textcolor{black}{NASA} Extragalactic Database, NED, is operated by the Jet Propulsion Laboratory, California Institute of Technology, under contract with the National Aeronautics and Space Administration.
We acknowledge the usage of the HyperLeda database (http://leda.univ-lyon1.fr).
Funding for the SDSS and SDSS-II has been provided by the Alfred P. Sloan Foundation, the Participating Institutions, the National Science Foundation, the U.S. Department of Energy, the National Aeronautics and Space Administration, the Japanese Monbukagakusho, the Max Planck Society, and the Higher Education Funding Council for England. The SDSS Web Site is http://www.sdss.org/. The SDSS is managed by the Astrophysical Research Consortium for the Participating Institutions. The Participating
– 13 –
Institutions are the American Museum of Natural History, Astrophysical Institute Potsdam, University of Basel, University of Cambridge, Case Western Reserve University, University of Chicago, Drexel University, Fermilab, the Institute for Advanced Study, the Japan Participation Group, Johns Hopkins University, the Joint Institute for Nuclear Astrophysics, the Kavli Institute for Particle Astrophysics and Cosmology, the Korean Scientist Group, the Chinese Academy of Sciences (LAMOST), Los Alamos National Laboratory, the Max-Planck-Institute for Astronomy (MPIA), the Max-Planck-Institute for Astrophysics (MPA), New Mexico State University, Ohio State University, University of Pittsburgh, University of Portsmouth, Princeton University, the United States Naval Observatory, and the University of Washington.  Based on observations made with the NASA Galaxy Evolution Explorer.
GALEX is operated for NASA by the California Institute of Technology under NASA contract NAS5-98034.}

\bibliographystyle{aa} 
\bibliography{biblio}

\begin{thebibliography}{38}
\expandafter\ifx\csname natexlab\endcsname\relax\def\natexlab#1{#1}\fi

\bibitem[{{Arp} \& {Sulentic}(1985)}]{arp85}
{Arp}, H. \& {Sulentic}, J.~W. 1985, ApJ, 291, 88

\bibitem[{{Baars} {et~al.}(1977){Baars}, {Genzel}, {Pauliny-Toth}, \&
  {Witzel}}]{baars77}
{Baars}, J.~W.~M., {Genzel}, R., {Pauliny-Toth}, I.~I.~K., \& {Witzel}, A.
  1977, A\&A, 61, 99

\bibitem[{{Bell} \& {de Jong}(2001)}]{Bell}
{Bell}, E.~F. \& {de Jong}, R.~S. 2001, \apj, 550, 212

\bibitem[{{Bergvall} \& {Ronnback}(1995)}]{bergvall95}
{Bergvall}, N. \& {Ronnback}, J. 1995, MNRAS, 273, 603

\bibitem[{{Bournaud} \& {Combes}(2002)}]{bournaud02}
{Bournaud}, F. \& {Combes}, F. 2002, A\&A, 392, 83

\bibitem[{{Bournaud} {et~al.}(2005){Bournaud}, {Combes}, {Jog}, \&
  {Puerari}}]{bournaud05}
{Bournaud}, F., {Combes}, F., {Jog}, C.~J., \& {Puerari}, I. 2005, A\&A, 438,
  507

\bibitem[{{Braun} {et~al.}(2003){Braun}, {Thilker}, \& {Walterbos}}]{braun03}
{Braun}, R., {Thilker}, D., \& {Walterbos}, R.~A.~M. 2003, A\&A, 406, 829

\bibitem[{{Condon} \& {Yin}(1990)}]{1990ApJ...357...97C}
{Condon}, J.~J. \& {Yin}, Q.~F. 1990, \apj, 357, 97

\bibitem[{{Conselice}(2003)}]{consel03}
{Conselice}, C.~J. 2003, ApJS, 147, 1

\bibitem[{{Durbala} {et~al.}(2008){Durbala}, {Sulentic}, {Buta}, \&
  {Verdes-Montenegro}}]{durbala08}
{Durbala}, A., {Sulentic}, J.~W., {Buta}, R., \& {Verdes-Montenegro}, L. 2008,
  MNRAS, 390, 881

\bibitem[{{Espada} {et~al.}(2005){Espada}, {Bosma}, {Verdes-Montenegro},
  {Athanassoula}, {Leon}, {Sulentic}, \& {Yun}}]{espada05}
{Espada}, D., {Bosma}, A., {Verdes-Montenegro}, L., {et~al.} 2005, \aap, 442,
  455

\bibitem[{{Espada} {et~al.}(2011{\natexlab{a}}){Espada}, {Mu{\~n}oz-Mateos},
  {Gil de Paz}, {Sabater}, {Boissier}, {Verley}, {Athanassoula}, {Bosma},
  {Leon}, {Verdes-Montenegro}, {Yun}, \& {Sulentic}}]{2011ApJ...736...20E}
{Espada}, D., {Mu{\~n}oz-Mateos}, J.~C., {Gil de Paz}, A., {et~al.}
  2011{\natexlab{a}}, \apj, 736, 20

\bibitem[{{Espada} {et~al.}(2011{\natexlab{b}}){Espada}, {Verdes-Montenegro},
  {Huchtmeier}, {Sulentic}, {Verley}, {Leon}, \& {Sabater}}]{espada11}
{Espada}, D., {Verdes-Montenegro}, L., {Huchtmeier}, W.~K., {et~al.}
  2011{\natexlab{b}}, A\&A, 532, A117

\bibitem[{{Fern{\'a}ndez Lorenzo} {et~al.}(2012){Fern{\'a}ndez Lorenzo},
  {Sulentic}, {Verdes-Montenegro}, {Ruiz}, {Sabater}, \&
  {S{\'a}nchez}}]{fernandez12}
{Fern{\'a}ndez Lorenzo}, M., {Sulentic}, J., {Verdes-Montenegro}, L., {et~al.}
  2012, A\&A, 540, 47

\bibitem[{{Haynes} \& {Giovanelli}(1984)}]{hayn84}
{Haynes}, M.~P. \& {Giovanelli}, R. 1984, AJ, 89, 758

\bibitem[{{Hern{\'a}ndez-Toledo} {et~al.}(2011){Hern{\'a}ndez-Toledo},
  {Cano-D{\'{\i}}az}, {Valenzuela}, {Puerari}, {Garc{\'{\i}}a-Barreto},
  {Moreno-D{\'{\i}}az}, \& {Bravo-Alfaro}}]{toledo2011}
{Hern{\'a}ndez-Toledo}, H.~M., {Cano-D{\'{\i}}az}, M., {Valenzuela}, O.,
  {et~al.} 2011, \aj, 142, 182

\bibitem[{{Hopkins} {et~al.}(2002){Hopkins}, {Schulte-Ladbeck}, \&
  {Drozdovsky}}]{2002AJ....124..862H}
{Hopkins}, A.~M., {Schulte-Ladbeck}, R.~E., \& {Drozdovsky}, I.~O. 2002, \aj,
  124, 862

\bibitem[{{James} {et~al.}(2004){James}, {Shane}, {Beckman}, {Cardwell},
  {Collins}, {Etherton}, {de Jong}, {Fathi}, {Knapen}, {Peletier}, {Percival},
  {Pollacco}, {Seigar}, {Stedman}, \& {Steele}}]{james04}
{James}, P.~A., {Shane}, N.~S., {Beckman}, J.~E., {et~al.} 2004, A\&A, 414, 23

\bibitem[{{Karachentseva}(1973)}]{kara73}
{Karachentseva}, V.~E. 1973, Astrofizicheskie Issledovaniia Izvestiya
  Spetsial'noj Astrofizicheskoj Observatorii, 8, 3

\bibitem[{{Leon} {et~al.}(2008){Leon}, {Verdes-Montenegro}, {Sabater},
  {Espada}, {Lisenfeld}, {Ballu}, {Sulentic}, {Verley}, {Bergond}, \&
  {Garc{\'{\i}}a}}]{2008A&A...485..475L}
{Leon}, S., {Verdes-Montenegro}, L., {Sabater}, J., {et~al.} 2008, \aap, 485,
  475

\bibitem[{{Lisenfeld} {et~al.}(2011){Lisenfeld}, {Espada}, {Verdes-Montenegro},
  {Kuno}, {Leon}, {Sabater}, {Sato}, {Sulentic}, {Verley}, \&
  {Yun}}]{2011A&A...534A.102L}
{Lisenfeld}, U., {Espada}, D., {Verdes-Montenegro}, L., {et~al.} 2011, \aap,
  534, A102

\bibitem[{{Lisenfeld} {et~al.}(2007){Lisenfeld}, {Verdes-Montenegro},
  {Sulentic}, {Leon}, {Espada}, {Bergond}, {Garc{\'{\i}}a}, {Sabater},
  {Santander-Vela}, \& {Verley}}]{lisen07}
{Lisenfeld}, U., {Verdes-Montenegro}, L., {Sulentic}, J., {et~al.} 2007, A\&A,
  462, 507

\bibitem[{{Pilyugin} {et~al.}(2004){Pilyugin}, {V{\'{\i}}lchez}, \&
  {Contini}}]{pilyu04}
{Pilyugin}, L.~S., {V{\'{\i}}lchez}, J.~M., \& {Contini}, T. 2004, A\&A, 425,
  849

\bibitem[{{Pohlen} \& {Trujillo}(2006)}]{pohlen06}
{Pohlen}, M. \& {Trujillo}, I. 2006, \aap, 454, 759

\bibitem[{{Portas} {et~al.}(2011){Portas}, {Scott}, {Brinks}, {Bosma},
  {Verdes-Montenegro}, {Heesen}, {Espada}, {Verley}, {Sulentic}, {Sengupta},
  {Athanassoula}, \& {Yun}}]{portas}
{Portas}, A., {Scott}, T.~C., {Brinks}, E., {et~al.} 2011, \apjl, 739, L27

\bibitem[{{Roberts} \& {Haynes}(1994)}]{roberts94}
{Roberts}, M.~S. \& {Haynes}, M.~P. 1994, ARA\&A, 32, 115

\bibitem[{{Sabater} {et~al.}(2008){Sabater}, {Leon}, {Verdes-Montenegro},
  {Lisenfeld}, {Sulentic}, \& {Verley}}]{2008A&A...486...73S}
{Sabater}, J., {Leon}, S., {Verdes-Montenegro}, L., {et~al.} 2008, \aap, 486,
  73

\bibitem[{{Sancisi} {et~al.}(2008){Sancisi}, {Fraternali}, {Oosterloo}, \& {van
  der Hulst}}]{sancisi08}
{Sancisi}, R., {Fraternali}, F., {Oosterloo}, T., \& {van der Hulst}, T. 2008,
  A\&AR, 15, 189

\bibitem[{{Struck}(1999)}]{struck99}
{Struck}, C. 1999, Physics Reports, 321, 1

\bibitem[{{Tifft} \& {Cocke}(1988)}]{tiff88}
{Tifft}, W.~G. \& {Cocke}, W.~J. 1988, ApJS, 67, 1

\bibitem[{{van Gorkom}(2004)}]{vgork04}
{van Gorkom}, J.~H. 2004, in Clusters of Galaxies: Probes of Cosmological
  Structure and Galaxy Evolution, ed. J.~S. {Mulchaey}, A.~{Dressler}, \&
  A.~{Oemler}, 305

\bibitem[{{Verdes-Montenegro} {et~al.}(2005){Verdes-Montenegro}, {Sulentic},
  {Lisenfeld}, {Leon}, {Espada}, {Garcia}, {Sabater}, \& {Verley}}]{vm05}
{Verdes-Montenegro}, L., {Sulentic}, J., {Lisenfeld}, U., {et~al.} 2005, A\&A,
  436, 443

\bibitem[{{Verley} {et~al.}(2007b){Verley}, {Leon}, {Verdes-Montenegro},
  {Combes}, {Sabater}, {Sulentic}, {Bergond}, {Espada}, {Garc{\'{\i}}a},
  {Lisenfeld}, \& {Odewahn}}]{verley07b}
{Verley}, S., {Leon}, S., {Verdes-Montenegro}, L., {et~al.} 2007b, A\&A, 472,
  121

\bibitem[{{Verley} {et~al.}(2007a){Verley}, {Odewahn}, {Verdes-Montenegro},
  {Leon}, {Combes}, {Sulentic}, {Bergond}, {Espada}, {Garc{\'{\i}}a},
  {Lisenfeld}, \& {Sabater}}]{verley07a}
{Verley}, S., {Odewahn}, S.~C., {Verdes-Montenegro}, L., {et~al.} 2007a, \aap,
  470, 505

\bibitem[{{Wong} {et~al.}(2006){Wong}, {Meurer}, {Bekki}, {Hanish},
  {Kennicutt}, {Bland-Hawthorn}, {Ryan-Weber}, {Koribalski}, {Kilborn},
  {Putman}, {Heiner}, {Webster}, {Allen}, {Dopita}, {Doyle}, {Drinkwater},
  {Ferguson}, {Freeman}, {Heckman}, {Hoopes}, {Knezek}, {Meyer}, {Oey},
  {Seibert}, {Smith}, {Staveley-Smith}, {Thilker}, {Werk}, \&
  {Zwaan}}]{2006MNRAS.370.1607W}
{Wong}, O.~I., {Meurer}, G.~R., {Bekki}, K., {et~al.} 2006, \mnras, 370, 1607

\bibitem[{{Woods} {et~al.}(2006){Woods}, {Geller}, \& {Barton}}]{woods06}
{Woods}, D.~F., {Geller}, M.~J., \& {Barton}, E.~J. 2006, AJ, 132, 197

\bibitem[{{York} {et~al.}(2000){York}, {Adelman}, {Anderson}, {Anderson},
  {Annis}, {Bahcall}, \& {Bakken}}]{york00}
{York}, D.~G., {Adelman}, J., {Anderson}, Jr., J.~E., {et~al.} 2000, AJ, 120,
  1579

\bibitem[{{Yun} {et~al.}(2001){Yun}, {Reddy}, \& {Condon}}]{yun01}
{Yun}, M.~S., {Reddy}, N.~A., \& {Condon}, J.~J. 2001, ApJ, 554, 803

\end{thebibliography}




\end{document}